# Two-Phase Model of the Polycrystalline Aggregate with Account for Grain-Boundary States under Quasi-Static Deformation

A.A. Reshetnyak[1,a)] and Yu.P. Sharkeev[1,2,b)]

[1]*Institute of Strength Physics and Materials Science SB RAS, 2/4, av. Akademicheskii, Tomsk, 634055, Russia*
[1,2]*National Research Tomsk Polytechnic University, 10, av. Lenin, Tomsk, 634011, Russia*

a)Corresponding author: reshet@ispms.tsc.ru
b)Sharkeev@ ispms.tsc.ru

**Abstract.** The recently suggested statistical theory of flow stress, including yield strength, for polycrystalline materials under quasi-static plastic deformation is developed in the framework of a two-phase model. Analytic and graphic forms of the generalized Hall—Petch relations are obtained for samples with BCC (α-phase Fe), FCC (Cu, Al, Ni) and HCP (α-Ti, Zr) crystalline lattices at T=300K with different values of grain-boundary (second) phase. The maximum of yield strength and respective extremal grain size of the samples are shifted by changing of the second phase. Temperature dependence in the range of 100-350K for yield strength (using the example of Al) revealed its increase for closely packed nano-crystalline samples with the growth of temperature. An enlargement of the second phase in a sample neutralizes this property.

## INTRODUCTION

The theory of flow stress for polycrystalline (PC) materials under quasi-static plastic deformation (PD), in the case of tensile strain, depending on the average size of crystallites (grains) $d$ in the range of $10^{-8}$ m - $10^{-2}$ m, suggested in [1,2] is based on a statistical model of mechanical energy distribution of each crystallite of single-mode isotropic PC material with respect to quasi-stationary levels. The largest energy level is chosen to be equal to the energy of maximal straight-line dislocation in the framework of disclination-dislocation deformation mechanism. The energy spectrum of each crystallite consists of equal-distant energy zones (realizing the most probable assembly of dislocations with a unique Burgers vector, b, starting from the zero-level energy of a crystallite without defects, $E_0$, up to the level with a maximal value of the atoms on a dislocation axis, $E_N$, $N=d/b$. Part (quantum) of the energy necessary for transition from a crystallite state to a neighboring state equals to the energy of a unit dislocation, ½Gb³, being almost precisely measurable with the activation energy of an atom in a material during the diffusion process.

A non-equilibrium deformation process was presented in [1,2] as a sequence of equilibrium deformation processes, with allowance for the smallness of relaxation time for the atoms of the crystal lattice of a grain in a new ground states as compared to a minimum duration between the acts of PD. On the segments of equilibrium at a given strain, ε, the emergence probabilities of possible defects at an elementary act of PD at the moment $t=\varepsilon/\dot{\varepsilon}$ for a deformation rate, $\dot{\varepsilon}$, are distributed (in accordance with the Large numbers law) according to Boltzmann, $P_n(\varepsilon) \sim A(\varepsilon) exp\{-\frac{n}{2} G b_\varepsilon^3 / NkT\}$, with an effective Burgers vector, $b_\varepsilon = b(1+\varepsilon)$: $b_\varepsilon(T') = b_\varepsilon(T)[1+\alpha_d(T'-T)]$ and shear modulus $G(T')=G(T)[1-\alpha_G(T'-T)]$ with linear temperature coefficients $\alpha_d$, $\alpha_G$ for T'>T being different for various temperature ranges in a given phase of a material.

The distribution of scalar density of dislocations [1], $\rho$, in each crystallite leads to a flow stress $\sigma(\varepsilon)$ according to the Taylor strain hardening mechanism, which contains at ε=0,002 normal [3] and abnormal (see, e.g., [4,5] and references therein) Hall—Petch (HP) relations for coarse-grained (CG) and nano-crystalline (NC) grains. Second, $\sigma(\varepsilon)$ has a maximum for the flow stress reached at extreme grain of the average value $d_0$ [1]:

$$d_0(\varepsilon, T) = b \frac{Gb^3(1+\varepsilon)^3}{2 \cdot 1{,}59363 \cdot kT}, \qquad (1)$$

which is estimated as $10^{-8}$-$10^{-7}$m and shifted to the region of larger grains with a decrease in temperature and increase in strain ε. It is demonstrated an appearance and evolution of cellular and cellular-mesh dislocation substructures, originated at the accumulation of strain in view of the curvature of a crystal lattice (CL) to be changed depending on $d$. The density $\rho$ obtained on a basis of mechanical arguments can be equivalently calculated with a quasi-particle interpretation for the energy quantum of PD, conditionally called a *dislocon* [1,2]. This object within the L.de Broglie corpuscular-wave dualism is nothing else then a composite quasi-particle consisting of acoustic phonons which surround the atoms at a broken bond between them in the CL. These phonons are not free and are joined in dislocons only for a non-elastic action in CL [2]. In the framework of the suggested statistical theory of flow stress, the coincidence of theoretical and experimental data is reached for yield strength, $\sigma_y$, extreme sizes $d_0$ for materials with BCC (α-Fe), FCC (Cu, Al, Ni) and HCP (α-Ti, Zr) CLs for T=300K within a one-phase approximation and without a grain boundary (GB) weak phase. The temperature dependence of strength characteristics was studied. It was shown, using the example of Al, that $\sigma_y$ increases when a decrease in T for all grains is more than $3d_0$ and then decreases in the NC region [2]. The stress-strain curves for α-Fe are constructed for the crystalline phase of α-Fe with account of the Backofen—Considere fracture criterion validity.

The one-phase model for a PC material was extended in [2] by including the un-hardening GB phase, up to two-phase model with integral flow stress:

$$\sigma_\Sigma(\varepsilon) = \left(1 - n\frac{b}{d}\right)\sigma_C(\varepsilon, d) + (n-m)\frac{b}{d}\sigma_{GB}(\varepsilon, d_{GB}) - m\frac{b}{d}\sigma_P(\varepsilon, d_P), m \leq n$$

$$\sigma_C(\varepsilon, d) = \sigma(\varepsilon) = \sigma_0(\varepsilon) + \alpha m \frac{Gb}{d}\sqrt{\frac{6\sqrt{2}}{\pi}m_0 \varepsilon M(0)}\left(e^{M(\varepsilon)\left[\frac{b}{d}\right]} - 1\right)^{-\frac{1}{2}}, \ M(\varepsilon) = Gb_\varepsilon^3/2kT, \quad (2)$$

for $\sigma_C(\varepsilon, d) = \sigma(\varepsilon)$ being the flow stress for the 1-st (solid) phase with basic grains of a sample of diameter, $d$, for $\sigma_{GB}(\varepsilon, d_{GB})$ and $\sigma_P(\varepsilon, d_P)$, the flow stresses having the same dependence as $\sigma_C(\varepsilon, d) = \sigma(\varepsilon, d)$, albeit for the grains and pores from the GB region of average sizes $d_{GB}$ and $d_P$, respectively, with some constant $n \sim 10^0 - 10^2$, taking account of the average distances between the grains and strongly depending on the preparation of GB states. For $n = m$, all regions of the 2-nd phase is filled by pores of various diameters. An input from the latter in $\sigma_\Sigma(\varepsilon)$ is negative conditionally, corresponding to an "anti-crystalline" behavior of pores with respect to which in submicrocrystalline (SMC) and NC samples one realizes a slippage of (group) grains. The value $m_0$ in (2) is the polyhedral parameter [1,2], which is determined for CG materials from the limit of the normal HP law for the flow stress (2) at $\varepsilon = 0{,}002$ by the relation with experimentally found Hall—Petch coefficient $k(\varepsilon)$ [1] in the approximation of one-phase model:

$$\sigma(\varepsilon)|_{d\gg b} = \sigma_0(\varepsilon) + k(\varepsilon)d^{-\frac{1}{2}}, \ k(\varepsilon) = \alpha m G\sqrt{\frac{6\sqrt{2}}{\pi}m_0 \varepsilon b \ M(0)/M(\varepsilon)} \ \Rightarrow \ m_0 = \frac{\pi}{6\sqrt{2}}\frac{k^2(\varepsilon)}{(\alpha m G)^2 \varepsilon b}\frac{M(\varepsilon)}{M_0}. \quad (3)$$

In this paper we develop a study of theoretical HP relations for PC materials: α-Fe, Cu, Al, Ni, α-Ti, Zr [1,2], by inclusion of the GB phase in the cases of small-angle and large-angle grain boundaries as well as for the constant average diameter of pores, $d_P = \bar{d}_P$, in the entire admissible range of grains $d$ from the crystallite phase with new values $d_{\Sigma 0}$ of the extreme grains and maxima $\sigma_{\Sigma m}$. Second, we investigate a modification of temperature dependence of the HP law for Al within the above realizations of weak phase of a PC aggregate.

## GENERALIZED HALL-PETCH LAW FOR α-Fe, Cu, Al, Ni, α-Ti, Zr

First, one should note the flow stress $\sigma_\Sigma(\varepsilon)$ (2) may be naturally extended for the case of dispersion hardening to $\sigma_{\Sigma dis}(\varepsilon) = (1 - U_{dis})\sigma_C(\varepsilon) + U_{dis}\sigma_{dis}(\varepsilon, d_{dis})$ by the "third phase" term: $\sigma_{dis}(\varepsilon, d_{dis})$ with the weight $0 \leq U_{dis} < 1$, for $d_{dis}$ being by average linear size of particles (from another compounds) which realize this hardening process. Here, the role of $\sigma_{dis}$ is analogous to one of $\sigma_{GB}(\varepsilon, d_{GB})$ for the second phase grains, but with proper shear modulus and Burgers vector $G_{dis}, b_{dis}$ now distributed inside of the first phase grains. For the CG and other PC samples with $d_{dis} \ll d$ the particles may provide the growth of the integral $\sigma_{\Sigma dis}(\varepsilon)$, in particular yield strength, when $Gb^3 < G_{dis}b_{dis}^3$. But for NC samples the hardening could take place for more complicated conditions: $d_0 < d_{dis} < d$ and for, possibly $Gb^3 < G_{dis}b_{dis}^3$, and un-hardening possibly for $d_{dis} < d_0 < d$ with some relation among unit dislocation energies $\frac{1}{2}Gb^3$, $\frac{1}{2}G_{dis}b_{dis}^3$ following to the results [1,2].

To determine the values of the constant $m_0$ (3) for two-phase model let us use the known experimental values for HP coefficient $k(0{,}002)$ for PC unimodal samples with BCC, FCC and HCP CL from the Table 1 with small-angle GBs, corresponding values for $\sigma_0$, $G$, lattice constants $a$ [6], Burgers vectors with the least possible lengths $b$, with respective most realizable sliding systems (given in the Table 2 [2]), constant of interaction for the dislocation $\alpha$ [4,5] and computed values of the least unit dislocations $E_d^{L_e}$, extreme grain sizes $d_0$, maximal differences of yield strength $\Delta\sigma_m$, $\Delta\sigma_{\Sigma m}$ in accordance with (5) in [1] and (2) for T=300K:

**TABLE 1**: Values $\sigma_0$, $\Delta\sigma_m = (\sigma_m - \sigma_0)$, $\Delta\sigma_{\Sigma m} = (\sigma_{\Sigma m} - \sigma_0)$, $E_d^{L_e}$, $k$, $m_0$, $\alpha$ for BCC, FCC and HCP polycrystalline metal samples with $d_0$, $b$, $G$, obtained with use of data from [1,4,6] at $\varepsilon = 0{,}002$ and $d_{\Sigma 0}$ obtained from the Fig. 1 for $(d_{Ps}, d_{Pg}) = (0{,}05, 0{,}2)*d$, that corresponds to average weights of the phases $(1-nb/d; nb/d)$ equal to $(0{,}952; 0{,}048)$ for small and $(0{,}833; 0{,}167)$ for large angles and constant size of porous $\bar{d}_P$ for each PC samples, when it exists. The lowest boundary $d_{LB}$ for the existence of the samples with $\bar{d}_P$ are estimated as $d_{LB}$(α-Fe; Cu; Al; Ni, α-Ti; Zr) ≈(21;13;13;22;22;23)nm.

| Type of CL | BCC | FCC | | | HCP | |
|---|---|---|---|---|---|---|
| **Material** | α-Fe | Cu | Al | Ni | α-Ti | Zr |
| $\sigma_0$, MPa | 170 (annealed) | 70 (anneal.); 380 (cold-worked) | 22 (anneal. 99,95%); 30 (99,5%) | 80 (annealed) | 100(~100%); 300 (99,6%) | 80-115 |

| $b$, nm | $\frac{\sqrt{3}}{2}a$ =0,248 | $a/\sqrt{2}$ =0,256 | $a/\sqrt{2}$ =0,286 | $a/\sqrt{2}$ =0.249 | $a$=0,295 | $a$=0,323 |
|---|---|---|---|---|---|---|
| $G$, GPa | 82,5 | 44 | 26,5 | 76 | 41,4 | 34 |
| $T$, K | 300 | 300 | 300 | 300 | 300 | 300 |
| $k$, MPa·m$^{1/2}$ | 0,55-0,65 ($10^{-5}$–$10^{-3}$ m) | 0,25 ($10^{-4}$–$10^{-3}$ m) | 0,15 ($10^{-4}$–$10^{-3}$ m) | 0,28 ($10^{-5}$–$10^{-3}$ m) | 0,38-0,43 ($10^{-5}$–$10^{-3}$ m) | 0.26 ($10^{-5}$–$10^{-3}$ m) |
| $\alpha$ | – | 0,38 | – | 0,35 | 0,97 | – |
| $E_d^{L_e} = \frac{1}{2}Gb^3$, eV | 3,93 | 1,28 | 1,96 | 3,72 | 3,33 | 3,57 |
| $m_0 \cdot \alpha^2$ | 3,66-5,11 | 2,57 | 2,28 | 1,11 | 5,83-7,47 | 3,69 |
| $d_0$, nm | 23,6 | 14,4 | 13,6 | 22,6 | 23,8 | 28,0 |
| $\Delta\sigma_m$, GPa | 2,29-2,69 | 1,31 | 0,83 | 1,18 | 1,58-1,79 | 0,99 |

| $d_{\Sigma 0}$, nm | | | | | | | | | | | | |
|---|---|---|---|---|---|---|---|---|---|---|---|---|
| $d_{Ps}$ | 23,5 | | 14,3 | | 13,5 | | 22,5 | | 23,5 | | 27,8 | |
| $d_{Pg}$ | 22,5 | | 13,5 | | 11,4 | | 18,9 | | 22,5 | | 25,0 | |
| $\bar{d}_P$ | ~120 | 24,0 | ~64 | 14,4 | ~64 | 13,6 | 114 | 22 | ~130 | 24,8 | 152 | 28 |

| $\Delta\sigma_{\Sigma m}$, GPa | | | | | | | | | | | | |
|---|---|---|---|---|---|---|---|---|---|---|---|---|
| $-\sigma_{Pms}$ | 2,19 | 0,11 | 1,28 | 0,06 | 0,79 | 0,04 | 1,14 | 0,05 | 1,51 | 0,08 | 0,95 | 0,05 |
| $-\sigma_{Pmg}$ | 1,85 | 0,45 | 1,08 | 0,22 | 0,69 | 0,13 | 0,98 | 0,2 | 1,28 | 0,26 | 0,85 | 0,16 |
| $-\bar{\sigma}_{Pm}$ | 1,02 | 1,19 | 0,57 | 0,69 | 0,34 | 0,43 | 0,53 | 0,6 | 0,73 | 0,83 | 0,47 | 0,54 |

The values for $k$ at $\varepsilon = 0,002$ are taken, e.g. for α-Fe, Cu, Ni from [4,5], for Al from [5], for Zr, α-Ti see [2] in the range for the grains shown in the brackets.

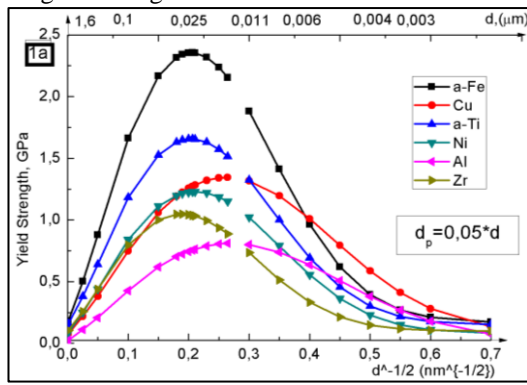

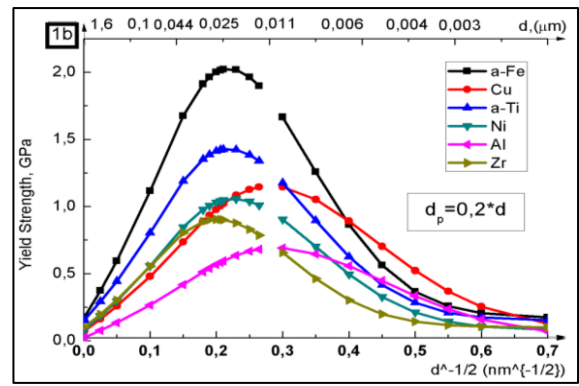

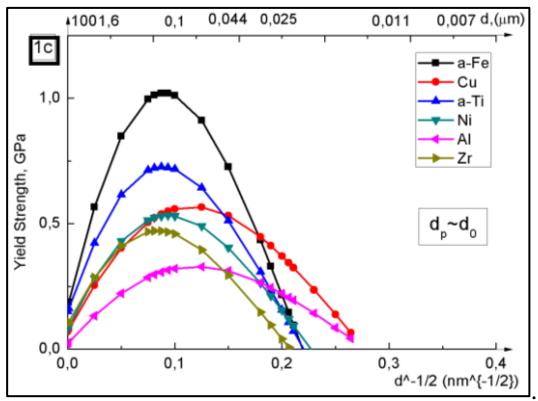

**FIGURE 1**. Graphical dependences for generalized Hall-Petch law (2) at $\varepsilon = 0,002$ with additional upper scale with size of the grains $d$ given in μm for small-angle (on the Fig. 1a, for $d_{Ps}$=0,05$d$), large-angle (on the Fig. 1b, for $d_{Pg}$=0,2$d$), for whole range of the crystallite size d, and for constant average porous $\bar{d}_P \approx d_0$ (on the Fig. 1c,) grain boundaries. Upper axis $d$ is changing within range (∞;0) with inverse quadratic scale and correspondence: (100; 1,6; 0,1; 0,044; 0.025; 0,011; 0,006; 0,004; 0,003) μm ↔ (0,005; 0,015; 0,1; 0,15; 0,2; 0,3; 0,41; 0,5; 0,57) nm$^{-1/2}$ for respective values on lower axis.

The least from possible values of the parameters $m_0(k)$ for α-Fe, α-Ti, values of $\sigma_0$ for annealed materials with maxima for $\sigma_y$ are calculated in the respective to the Table 1 extreme grain sizes $d_0$

## TEMPERATURE DEPENDENCE OF YIELD STRENGTH AND EXTREME GRAIN SIZES FOR Al

Because of with a growth of the temperature the value of the shear modulus $G(T)$ (as well as $\sigma_0(T)$) decreases, whereas the linear parameters $b$, $d$ are increased with the same true linear coefficient of the temperature expansion $\alpha_d$ [6] (for BCC and FCC materials, see as well [7]), then the extreme grain size $d_0(\varepsilon,T)$ is shifted in to the region of smaller grains: $d_0(\varepsilon,T) > d_0(\varepsilon, T')$ for $T > T'$ (within the same phase of the material [1,2]) according to the rule:

$$d_0(\varepsilon, T') = b_\varepsilon(T') \frac{\frac{1}{2}G(T')[b_\varepsilon(T')]^3}{1{,}59363 \cdot kT} = d_0(\varepsilon, T)g(\alpha_G, \alpha_d, T, T'),$$
$$g(\alpha_G, \alpha_d, T, T') = \left(\frac{b_\varepsilon(T')}{b_\varepsilon(T)}\right)^4 \frac{G(T')T}{G(T)T'} = (1 + \alpha_d(T' - T))^4 (1 - \alpha_G(T' - T))\frac{T}{T'}, \quad (4)$$
$$\text{for } b_\varepsilon(T') = b_\varepsilon(T)(1 + \alpha_d(T' - T)); \quad G(T') = (1 - \alpha_G(T' - T))G(T)$$

with linear temperature coefficient for the shear modulus $\alpha_G$, e.g. for Al: $\alpha_G = 5{,}2 \cdot 10^{-4} K^{-1}$ being approximately constant within temperature range [250K,300K]. It follows from (4) that for varying of $T$ in a small range the value $d_0(\varepsilon,T)$ is multiplicatively changed with the factor $g(\alpha_G, \alpha_d, T, T')$. At the same time, with accommodation of PD the value of $d_0(\varepsilon,T)$ is shifted to the area of larger grains: $d_0(\varepsilon_1,T) > d_0(\varepsilon_2,T)$ for $\varepsilon_1 > \varepsilon_2$. Both behaviors of $\sigma(\varepsilon)$ and $\sigma_m(\varepsilon)$ for a monotonic change of the temperature is added from T-behavior of the crystal and dislocation substructures of both crystalline and weak phases of the sample according to (2), (4). The results of the theoretical study of the T-dependence for the Hall-Petch law for *Al* samples for large-angle GB and for the constant size of porous from the 2-nd phase in the whole range of diameters $d$ of the 1-st phase crystallites in comparison with pure crystallite phase realization [2] are given in the Table 2 and Fig. 2.

**TABLE 2** Values $G$, $\sigma_0$, $d_0$, $\Delta\sigma_m$ for PC *Al* samples with $d_0$, $b$, $G$ for $T$=300K are taken from the Table 1, Fig. 1 and Ref.[2] at $\varepsilon$=0,002 and $d_{\Sigma 0}$, $\Delta\sigma_{\Sigma m} = (\sigma_{\Sigma m} - \sigma_0)$ obtained from the Fig. 2 for $d_{Pg}$=0,25*d*, that corresponds to average weights of the phases (1-*nb/d; nb/d*)=(0,8; 0,2) without account for the weak phase grains for large-angle GB and constant size of porous $\bar{d}_P$ PC *Al* samples, when it exists. The boundary of existence $d_{LB}$ is estimated by 11nm

| Al  T,K | G, GPa | $\sigma_0$, MPa | $d_0$, nm | $(\sigma_m-\sigma_0)$, GPa | $d_{\Sigma 0}$, nm; ($d_{Pg}$=0,25d) | $\Delta\sigma_{\Sigma m}$, GPa | $d_{\Sigma 0}$, nm; ($\bar{d}_P$=13,6nm) | $\Delta\sigma_{\Sigma m}$, GPa |
|---|---|---|---|---|---|---|---|---|
| 350 | 25,8 | 21 | 11,7 | 0,85 | 8,5 | 0,65 | 70,0 | 0,29 |
| 300 | 26,5 | 22 | 13,6 | 0,81 | 11,1 | 0,62 | 70,0 | 0,31 |
| 250 | 27,4 | 23 | 16,3 | 0,74 | 13,5 | 0,58 | 70,0 | 0,32 |
| 200 | 28,1 | 23,5 | 20,4 | 0,67 | 16,0 | 0,52 | 70,0 | 0,34 |
| 150 | 28,8 | 24 | 27,2 | 0,59 | 23,5 | 0,46 | 70,0 | 0,36 |

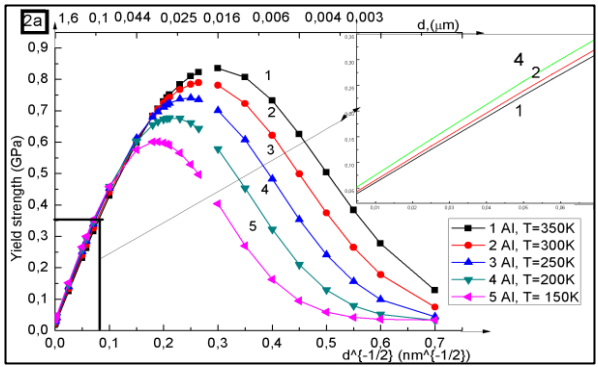
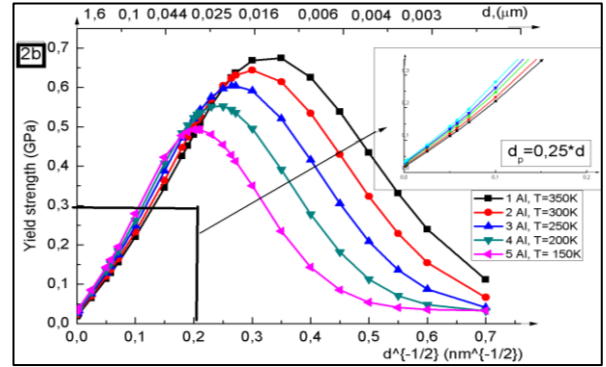
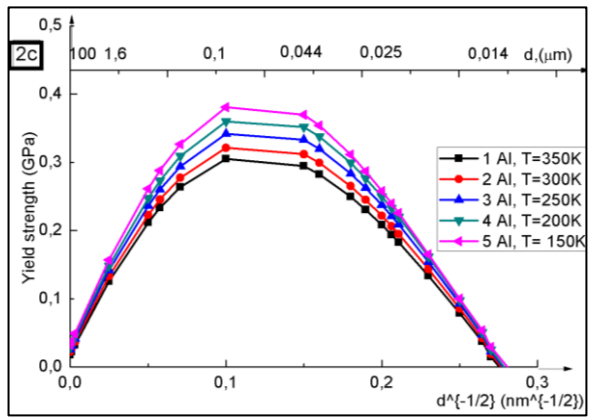

**FIGURE 2** Graphical dependences for generalized Hall-Petch law (2) at $\varepsilon = 0{,}002$ for Al at $T$=350; 300; 250; 200; 150$K$ for pure crystalline phase (on the Fig. 2a, [2]), for two-phase model with large-angle grain boundaries (on the Fig. 2b, for $d_p$=0,25*$d$) and for constant porous diameter $d_p$=$d_0$(300)=13,6 nm (on Fig. 2c,). On the inputs the part of the dependences in coarse- and fine-grained regions are shown (extracted as the rectangles) for the values of curves $T$=350K; 300K; 200K with normal HP law validity on the Fig. 2a and for the whole set of temperatures on the Fig. 2b. The straights (3), (5) for $T$=250K; 150K are situated on the input of the Fig. 2a, respectively between the straights (2), (4) and above the straight (4). The auxiliary upper horizontal axis, which enumerates direct linear size $d$ of the grains, plays the same role as it is described on the Fig.1a, 1b,1c.

## SUMMARY

From the Table 1 and the Fig. 1 data for the Hall-Petch law at T=300K for homogeneous unimodal PC samples of α-Fe, Cu, Al, Ni, α-Ti, Zr within the two-phase model described by (2), it follows that the extreme sizes $d_{\Sigma 0}$ for maximum $\Delta\sigma_{\Sigma m}$ of yield strength both decrease in comparison with $d_0$, $\Delta\sigma_m$ calculated according to (1), (2), within a one phase approximation model for d-independent weights of the 1-st and 2-nd phases: $d_{\Sigma 0}<d_0$, $\Delta\sigma_{\Sigma m}<\Delta\sigma_m$. For small-angle GB the difference $\Delta\sigma_{\Sigma m}$ is larger than that for a large-angle GB. For a constant size of pores $d_P \sim d_0$, the weights of the 1-st and 2-nd phases become *d*-dependent, transforming PC samples (with a decreasing in *d*) from PC aggregates with small-angle GB to one large-angle GB up to the impossibility of the NC region of their existence, due to $\sigma_\Sigma<0$. The extreme size of grains for such samples are shifted to easily reached values in experiments, $d_{\Sigma 0}$(α-Fe, Cu, Al, Ni, α-Ti, Zr)~(120;64;64;114;130;152)nm, from the SMC region with a decrease in $\Delta\sigma_{\Sigma m}$: (≤1GPa). In turn, the study of temperature dependence for PC *Al* samples within the range of [150; 350]K shows from the analysis of Table 2 and Fig. 2 that the temperature-size effect discovered within the one-phase model approximation in [2], should be valid for samples with a d-independent small- and large-angle GB, leading to an increasing in $\Delta\sigma_{\Sigma m}$ and $\sigma_\Sigma$ for all $d<d_{\Sigma 1}\sim 3d_{\Sigma 0}$ [2] with a growth of temperature. For grains with $d>d_{\Sigma 1}$, the behavior of $\sigma_\Sigma$ is opposite (and is therefore usual for coarse- and fine-grained PC aggregates). The presence of the 2-nd phase (with its *d*-independent weight) decreases the extreme grains size $d_{\Sigma 0}$(T) in comparison with $d_0$(T). However, the 2-nd phase with *d*-dependent weights of the 1-st and 2-nd phases with constant pores $d_p=d_0(300)=$ 13,6 nm leads to the disappearance of this effect in such PC samples, revealing the decreasing of $\sigma_\Sigma$ with a growth of temperature in the whole range of grain sizes *d* with a stabilizing of extreme sizes $d_{\Sigma 0}$ for the PC *Al* aggregate at 70 nm.

From the above theoretical study follows the necessity of an experimental check of the predicted temperature-size effect. Namely, one should verify an increase in the extreme size of the grain $d_{\Sigma 0}(\varepsilon,T)$ with a decreasing temperature and separately with a growth of accumulated strain ε, and with a simultaneous decrease for a small-angle GB of the maximum $\sigma_\Sigma(\varepsilon)$. The mentioned phenomena are characteristic of NC and SMC PC samples. The suggested theoretical model has obvious perspectives of application to new PC materials in aircraft and cosmic technique, and is experimentally considered using samples with α-Ti alloy, BT1-0, in ultrafine-grained PC samples [8]. Having an NC coating with closely packed crystallites of PC materials on cosmic devices, one can provide its high-temperature strength when moving in dense atmosphere layers because the theory predicts an increase in the strength characteristics of the coating with a growth of temperature under the condition of its phase-state invariance. The model has permitted (on a basis of simple physical concepts and an application of statistical physics and probability theory technique) to obtain an analytic expression for the "σ-ε" dependence of PC materials in the whole range of grain sizes, values of accumulated strain, temperature, correctly corresponding to the experimental data One expected its further development for the case of many dislocation ensembles and multimodal PC samples*.

## ACKNOWLEDGMENTS

Authors are grateful to the referee's critical comments which have permitted to significantly improve the table and graphic presentations. A.A.R. sincerely acknowledges a help of the medical personal of the neurosurgical department of the Tomsk regional clinical hospital, where the significant part of the paper was done during his treatment. The work has been supported under the Program of fundamental research at the State Academy of Sciences for 2013-2020.

## REFERENCES

1. A.A.Reshetnyak, Statistical approach to flow stress and generalized Hall-Petch law for polycrystalline materials under plastic deformations, submitted in Russ. Phys. Journal (2018), Arxiv:1803.08247[cond-mat.mtr-sci].
2. A.A.Reshetnyak, Peculiriaties of temperature dependence for generalized Hall-Petch Law and two-phase model for deformable polycrystalline materials, submitted in Russ. Phys. Journal (2018). Arxiv:1805.08623[cond-mat.mtr-sci].
3. E.O. Hall. Proc. Roy. Soc. B **64**, 474 (1951); N.J. Petch. J. Iron Steel Inst. **174**, 25 (1953).
4. A.M. Glezer, E.V. Kozlov, N.A. Koneva et al. Plastic Deformation of Nanostructured Materials – CRC Press, 2017, 334p.
5. S.A. Firstov, Yu.F. Lugovskoi and B.A. Movchan, Structure, strength and fatigue resistance of microcrystalline and micro-layer materials. – Kiev: Naukova Dumka, 2016, 171p. (in Russian).
6. Physical Quantities. Guide. Edited I.S. Grigoriev, I.Z Meilihov. Moscow:Energoatomizdat–1991.–1232p [in Russian].
7. V.E.Panin and R.W.Armstrong, Phys. Mesomechanics, **19**, 35 (2016).
8. I.A. Kurzina, A.Yu.Eroshenko, Yu.P. Sharkeev et al., Materialovedenie. **5**, 49, (2010).